\documentclass[nofootinbib,twocolumn,aps,superscriptaddress,preprintnumbers]{myRevtex4}

\usepackage{graphicx}
\usepackage{color}
\usepackage{xspace}
\usepackage{refmerge}
\usepackage{natbib}   
\usepackage{tabularx}
\definecolor{darkblue1}{rgb}{0,0,.2}
\definecolor{darkblue}{rgb}{0,0,.2}
\definecolor{darkred}{rgb}{0.5,0,0}
\pagecolor{white} 
\color{black}     

\bibstyle{plain}
\mathchardef\Upsilon="7107
\def\Y#1S{\ensuremath{\Upsilon{(#1S)}}\xspace}

\newcommand{\mc}{\multicolumn}
\newcommand{\Tau}{\ensuremath{\tau}\xspace}
\newcommand{\nut}{\ensuremath{\nu_\tau}\xspace}

\newcommand{\BR}{\ensuremath{{\cal B}}\xspace}
\newcommand{\tautopipiz}{\ensuremath{\tau^-\to\pim\piz\nu_{\tau}}\xspace}
\newcommand{\tautoppz}{\tautopipiz}
\newcommand{\Btau}{\ensuremath{\BR_{\pipiz}}\xspace}

\newcommand{\BCVCppz}{\ensuremath{\BR^{\rm CVC}_{\pipiz}}\xspace}

\newcommand{\pip}{\ensuremath{\pi^+}\xspace}
\newcommand{\pim}{\ensuremath{\pi^-}\xspace}
\newcommand{\piz}{\ensuremath{\pi^0}\xspace}

\newcommand{\ee}{\ensuremath{e^+e^-}\xspace}

\newcommand{\pp}{\ensuremath{\pi^+\pi^-}\xspace}
\newcommand{\pipiz}{\ensuremath{\pi^-\pi^0}\xspace}
\newcommand{\ppz}{\ensuremath{\pi^-\pi^0}\xspace}

\newcommand{\Sew}{\ensuremath{S_{\rm EW}}\xspace}
\newcommand{\Gem}{\ensuremath{G_{\rm EM}}\xspace}
\newcommand{\tev}{\ensuremath{\mathrm{\,Te\kern -0.1em V}}\xspace}
\newcommand{\gev}{\ensuremath{\mathrm{\,Ge\kern -0.1em V}}\xspace}
\newcommand{\mev}{\ensuremath{\mathrm{\,Me\kern -0.1em V}}\xspace}
\newcommand{\kev}{\ensuremath{\mathrm{\,ke\kern -0.1em V}}\xspace}
\newcommand{\ev}{\ensuremath{\mathrm{\,e\kern -0.1em V}}\xspace}
\newcommand{\gevc}{\ensuremath{{\mathrm{\,Ge\kern -0.1em V\!/}c}}\xspace}
\newcommand{\mevc}{\ensuremath{{\mathrm{\,Me\kern -0.1em V\!/}c}}\xspace}
\newcommand{\gevcc}{\ensuremath{{\mathrm{\,Ge\kern -0.1em V\!/}c^2}}\xspace}
\newcommand{\mevcc}{\ensuremath{{\mathrm{\,Me\kern -0.1em V\!/}c^2}}\xspace}
\newcommand{\eetopp}{\ensuremath{\ee\to\pp}\xspace}

\newcommand{\beq}{\begin{equation}}
\newcommand{\eeq}{\end{equation}}
\newcommand{\beqn}{\begin{eqnarray}}
\newcommand{\eeqn}{\end{eqnarray}}
\newcommand{\beqns}{\begin{eqnarray*}}
\newcommand{\eeqns}{\end{eqnarray*}}
\newcommand{\bitm}{\begin{itemize}}
\newcommand{\eitm}{\end{itemize}}

\renewcommand{\arraystretch}{1.25}

\newcommand{\amuhadLO}{\ensuremath{a_\mu^{\rm had,LO}}\xspace}

\newcommand{\amuSM}{\ensuremath{a_\mu^{\rm SM}}\xspace}
\newcommand{\amuExp}{\ensuremath{a_\mu^{\rm exp}}\xspace}
\newcommand\eg{{\it e.g.}}

\newcommand\cf{{\em cf.}}
\newcommand{\ea}{{\em et al.}}
\newcommand\rs{\raisebox{1.5ex}[-1.5ex]}
\def\@citex[#1]#2{\if@filesw\immediate\write\@auxout{\string\citation{#2}}\fi
  \@tempcnta\z@\@tempcntb\m@ne\def\@citea{}\@cite{\@for\@citeb:=#2\do
    {\@ifundefined
       {b@\@citeb}{\@citeo\@tempcntb\m@ne\@citea
        \def\@citea{,\penalty\@m\ }{\bf ?}\@warning
       {Citation `\@citeb' on page \thepage \space undefined}}%
    {\setbox\z@\hbox{\global\@tempcntc0\csname b@\@citeb\endcsname\relax}%
     \ifnum\@tempcntc=\z@ \@citeo\@tempcntb\m@ne
       \@citea\def\@citea{,\penalty\@m}
       \hbox{\csname b@\@citeb\endcsname}%
     \else
      \advance\@tempcntb\@ne
      \ifnum\@tempcntb=\@tempcntc
      \else\advance\@tempcntb\m@ne\@citeo
      \@tempcnta\@tempcntc\@tempcntb\@tempcntc\fi\fi}}\@citeo}{#1}}

\def\@citeo{\ifnum\@tempcnta>\@tempcntb\else\@citea
  \def\@citea{,\penalty\@m}%
  \ifnum\@tempcnta=\@tempcntb\the\@tempcnta\else
   {\advance\@tempcnta\@ne\ifnum\@tempcnta=\@tempcntb \else
\def\@citea{--}\fi
    \advance\@tempcnta\m@ne\the\@tempcnta\@citea\the\@tempcntb}\fi\fi}
\newenvironment{myquote}
               {\list{}{\leftmargin0cm\indent}%
                \item\relax}
               {\endlist}
\newcommand\allFontSize{\footnotesize}
\newcommand\detailsSize{\allFontSize}
\newenvironment{details}%
{\begin{myquote}\detailsSize}{\end{myquote}}

\begin{document}

\preprint{\vbox{\hbox{BIHEP-TH-09-01, CERN-OPEN-2009-007, LAL 09-50}}}

\title{\boldmath The Discrepancy Between \Tau and \ee Spectral Functions Revisited and 
the \\
Consequences for the Muon Magnetic Anomaly}

\affiliation{Institute of High Energy Physics, Chinese Academy of Sciences, 
Beijing}
\affiliation{CERN, CH--1211, Geneva 23, Switzerland}
\affiliation{Laboratoire de l'Acc{\'e}l{\'e}rateur Lin{\'e}aire,
             IN2P3/CNRS, Universit\'e Paris-Sud 11, Orsay}
\affiliation{Departamento de F\'isica, Cinvestav, Apartado Postal 14-740, 
07000 M\'exico D.F., M\'exico}
\affiliation{Instituto de F\'isica, UNAM, A.P. 20-364, 01000 M\'exico D.F., 
M\'exico}

\author{M.~Davier}
\affiliation{Laboratoire de l'Acc{\'e}l{\'e}rateur Lin{\'e}aire,
             IN2P3/CNRS, Universit\'e Paris-Sud 11, Orsay}
\author{A.~Hoecker}
\affiliation{CERN, CH--1211, Geneva 23, Switzerland}
\author{G.~L\'opez Castro}
\affiliation{Departamento de F\'isica, Cinvestav, Apartado Postal 14-740, 
07000 M\'exico D.F., M\'exico}
\author{B.~Malaescu}
\affiliation{Laboratoire de l'Acc{\'e}l{\'e}rateur Lin{\'e}aire,
             IN2P3/CNRS, Universit\'e Paris-Sud 11, Orsay}
\author{X.H.~Mo}
\affiliation{Institute of High Energy Physics, Chinese Academy of Sciences, 
Beijing}
\author{G.~Toledo S\'anchez}
\affiliation{Instituto de F\'isica, UNAM, A.P. 20-364, 01000 M\'exico D.F., 
M\'exico}
\author{P.~Wang}
\author{C.Z.~Yuan}
\affiliation{Institute of High Energy Physics, Chinese Academy of Sciences, 
Beijing}
\author{Z.~Zhang}
\affiliation{Laboratoire de l'Acc{\'e}l{\'e}rateur Lin{\'e}aire,
             IN2P3/CNRS, Universit\'e Paris-Sud 11, Orsay}

\date{\today}

\begin{abstract}

We revisit the procedure for comparing the $\pi\pi$ spectral function
measured in \Tau decays to that obtained in \ee annihilation.
We re-examine the isospin-breaking corrections using new experimental and 
theoretical input, and find improved agreement between the \tautopipiz 
branching fraction measurement and its prediction using the
isospin-breaking-corrected \eetopp spectral function, though not resolving 
all discrepancies.
We recompute the lowest order hadronic contributions to the muon $g-2$ 
using \ee and \Tau data with the new corrections, and find a reduced 
difference between the two evaluations.
The new tau-based estimate of the muon magnetic anomaly is found to be 
$1.9$ standard deviations lower than the direct measurement.

\end{abstract}

\maketitle

\section{Introduction}

Spectral functions determined from the cross sections of \ee
annihilation to hadrons are fundamental quantities describing the
production of hadrons from the strong interaction vacuum. They are 
especially useful at low energy where perturbative QCD fails to describe
the data. Spectral functions play a crucial role in calculations of hadronic 
vacuum polarisation (VP) contributions to observables such as the effective 
electromagnetic coupling at the $Z^0$ mass, and the muon anomalous 
magnetic moment. The latter quantity requires good knowledge of the low 
energy spectral function dominated by the \pp channel.

During the last decade, measurements of the \pp spectral function 
with percent accuracy became available~\cite{cmd2,cmd2new,snd,kloe08},
superseding older and less precise data. The former lack 
of precision data inspired the search for an alternative. 
It was found~\cite{adh} in form of accurate \tautopipiz spectral 
functions~\cite{aleph_old,aleph_new,cleo,opal},
transferred from the charged to the neutral state using isospin symmetry.
With the increasing \eetopp experimental precision, which today is on a level 
with the \Tau data, systematic discrepancies in shape and normalisation of 
the spectral functions were observed between the two 
systems~\cite{dehz02,dehz03}. 
It was found that, when computing the hadronic VP contribution 
to the muon magnetic anomaly using the \Tau instead of the \ee data for the 
$2\pi$ and $4\pi$ channels, the observed deviation with the experimental 
value~\cite{bennett} would reduce from 3.3 times 
the combined experimental and estimated theoretical error to less than 
1~\cite{md_tau06}. 

In this paper, we include recent \tautopipiz data from the Belle experiment~\cite{belle}, 
and revisit all isospin-breaking corrections in this channel taking advantage of more accurate 
data and new theoretical investigations.

\begin{figure*}[htb]
\begin{center}
\includegraphics[width=89mm]{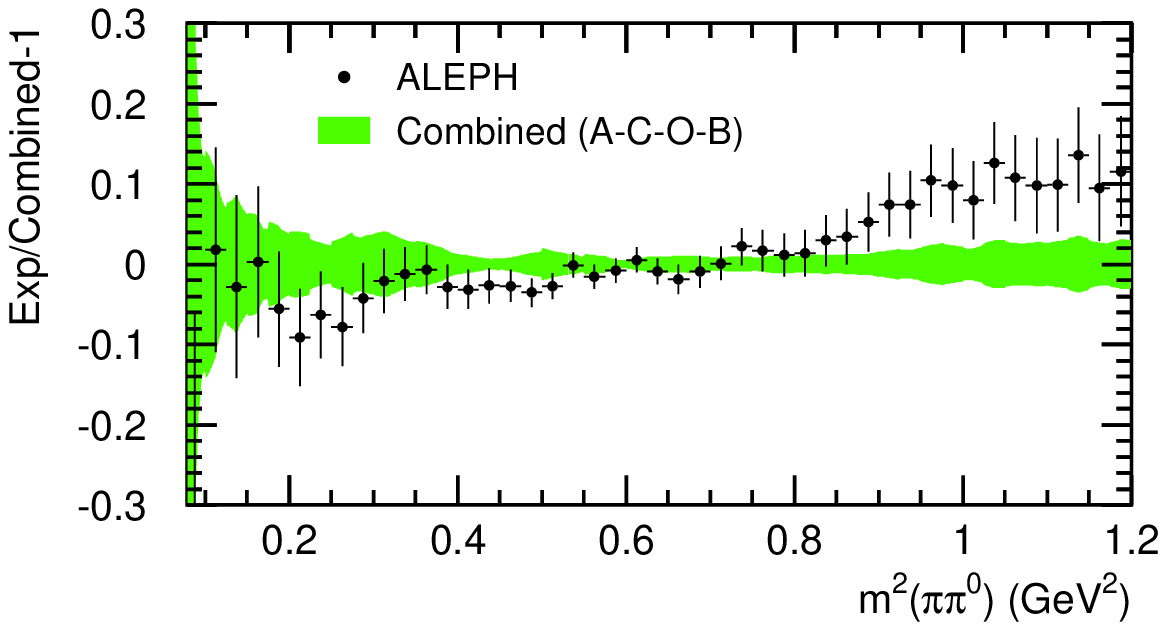}
\includegraphics[width=89mm]{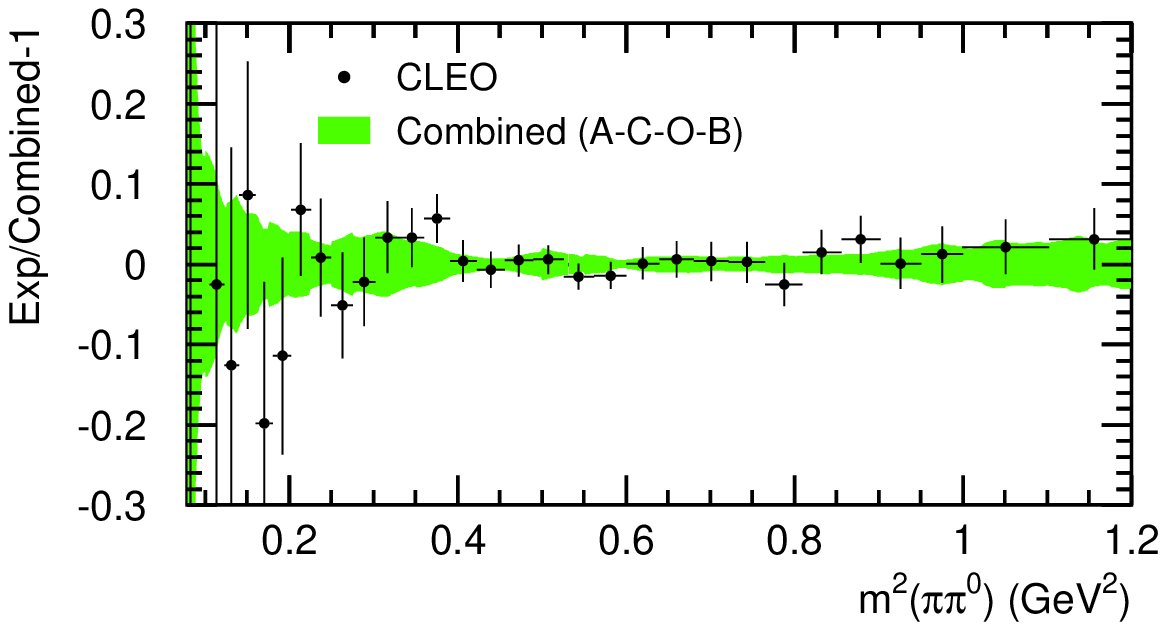} \\
\includegraphics[width=89mm]{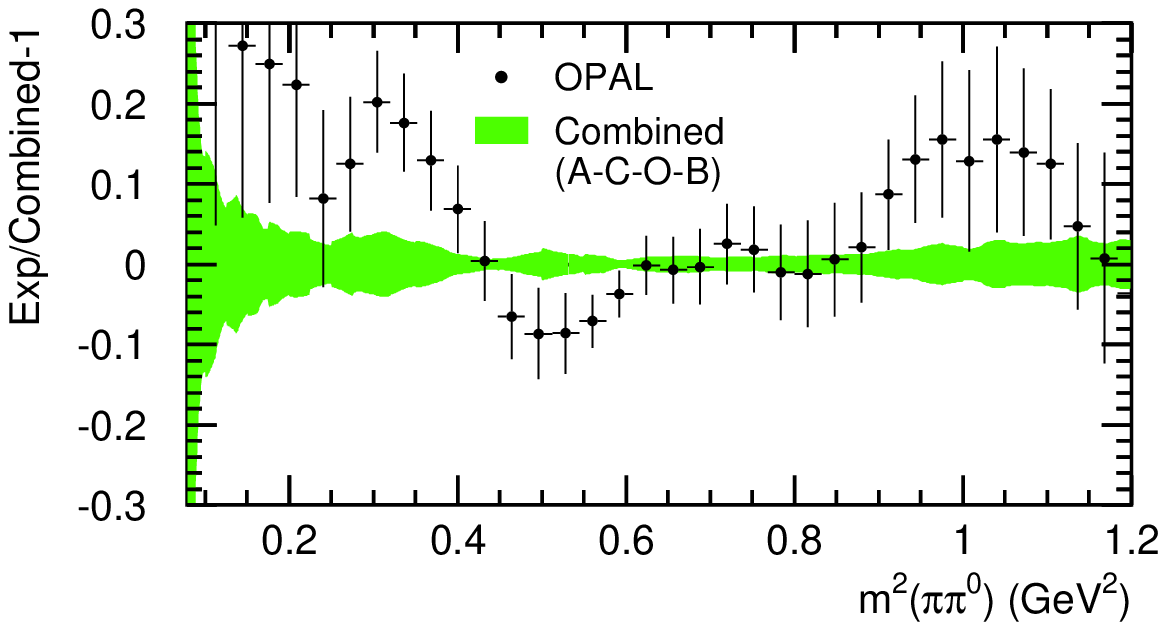} 
\includegraphics[width=89mm]{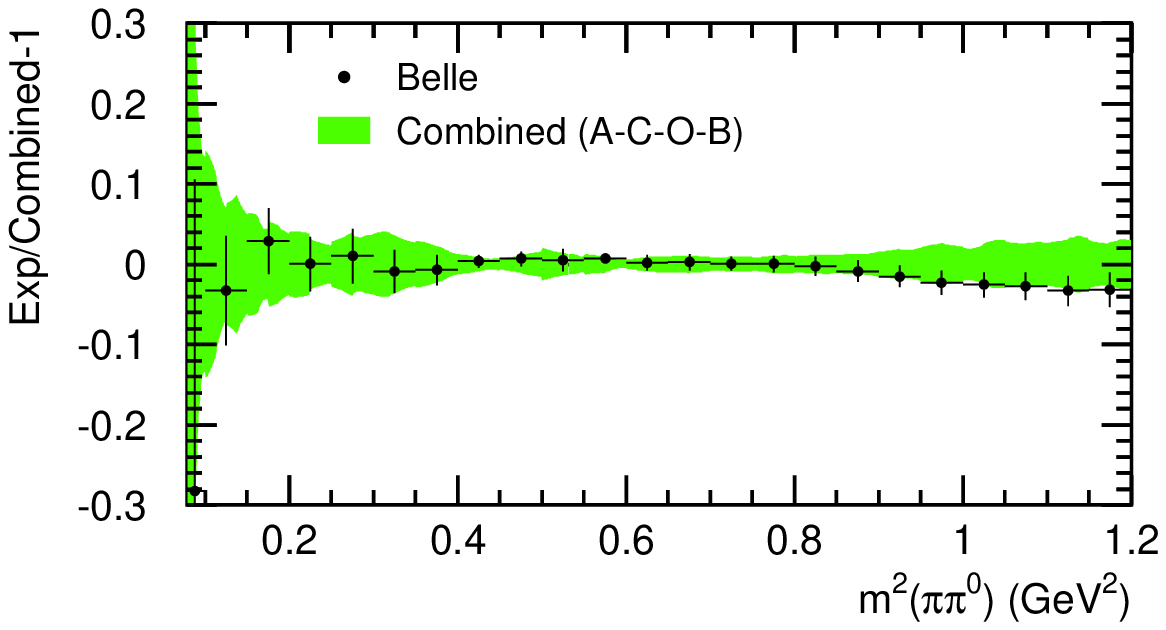}
\end{center}
\vspace{-0.3cm}
\caption{Relative comparison between the \tautopipiz invariant mass-squared 
measurements from ALEPH, CLEO, OPAL, Belle (data points) and 
the combined result (shaded band).}
\label{fig:tau_com}
\end{figure*}

\section{Tau data}

The $\tau$-based $a_\mu$ evaluation in~\cite{dehz02,dehz03} used 
the $\tau$ spectral functions measured by the ALEPH~\cite{aleph_new}, 
CLEO~\cite{cleo} and OPAL~\cite{opal} experiments for the dominant hadronic 
decay mode $\tau^-\to \pi^-\pi^0\nu_\tau$. We include here a high-statistics 
measurement of the same decay mode performed by Belle~\cite{belle}.
Rather different experimental conditions are met at the $Z$ centre-of-mass 
energy (ALEPH, OPAL) and at the $\Upsilon(4S)$ resonance (CLEO, Belle).
At LEP the $\tau^+\tau^-$ events can be selected with high efficiency ($>90\%$)
and small non-$\tau$ background ($<1\%$), thus ensuring little bias in the 
efficiency determination. The situation is not as favourable at low energy:
because the dominant hadronic cross section has a smaller particle
multiplicity, it is more likely to pollute the $\tau$ sample and strong
cuts must be applied, resulting in smaller selection efficiency with larger
relative uncertainty. On the other 
hand, the $\Upsilon(4S)$ machines outperform LEP in statistics for $\tau$-pair 
production: Belle's analysis contains 5.4 million 
$\tau^- \to h^-\pi^0\nu_\tau$ candidates ($72.2\,{\rm fb}^{-1}$ 
integrated luminosity), compared to $81$ thousand 
candidates used by ALEPH (including the full LEP statistics accumulated on the 
$Z$ pole). Moreover, CLEO and Belle have an advantage for the $\tau$ final 
state reconstruction since particles are more separated in space. 
The LEP detectors
have to cope with collimated $\tau$ decay products and the granularity of
the detectors, particularly the calorimeters, plays a crucial role.
One can therefore consider ALEPH/OPAL and CLEO/Belle data to be approximately 
uncorrelated as far as experimental procedures are concerned.\footnote
{
   Experimental correlations are introduced by common systematic errors in 
   the Monte Carlo simulation used. All experiments employ the same tau decay 
   and radiative corrections libraries, which are used for the correction 
   of feed-through from non-$h^-\pi^0$ final states, as well as for the 
   determination of the acceptance and efficiency after applying the selection 
   requirements. ALEPH and OPAL use however data-driven spectral functions for 
   the feed-through corrections, so that the resulting correlations should 
   be small. They are hence neglected in the average. 
} 
These four data sets are combined to provide the most precise $\tau$ spectral
function, using the newly developed software package
HVPTools~\cite{hvptools}. It transforms the original $\tau$ data and
associated statistical and systematic covariance
matrices into fine-grained energy bins ($1\,{\rm MeV}$).
In the combination, when the $\chi^2$ value of a bin-wise average exceeds 
the number of degrees of freedom ($n_{\rm dof}$), the error in the averaged 
bin is rescaled by $\sqrt{\chi^2/n_{\rm dof}}$ to account for
inconsistencies, which occur because most experiments are dominated by
systematic uncertainties.
Figure~\ref{fig:tau_com} shows the relative comparison of 
the combined data with those of each experiment.\footnote
{
   The total systematic uncertainty from Belle is derived from the correlated
   mass spectra containing revised information with respect to Table V of 
   Ref.~\cite{belle}, which has been provided to us by Belle~\cite{Hayashii}.
   They differ mostly at the low mass region below the $\rho$ mass peak.
}

The branching fraction $\BR_{\pi\pi^0}$ for 
$\tau^- \to\pi^-\pi^0\nu_\tau$ is obtained from the measured decay channel 
$\tau^- \to h^-\pi^0\nu_\tau$ ($\BR_{h\pi^0}$) 
by subtracting the non-$\pi$ contribution from the generic charged hadron 
mode ($h^-$).
The average $\BR_{h\pi^0}$ value from these experiments and the
two other LEP experiments L3~\cite{l3} and DELPHI~\cite{delphi} is
$(25.847\pm 0.101)\%$. Subtracting from this the current world average value
$(0.428\pm 0.015)\%$ for $\tau^- \to K^-\pi^0\nu_\tau$~\cite{pdg08},
gives $\BR_{\pi\pi^0}=(25.42\pm 0.10)\%$, which is the result 
used in the following.

\section{Isospin-breaking corrections}\label{sec:ib}

\begin{figure*}[htb]
\includegraphics[width=87mm]{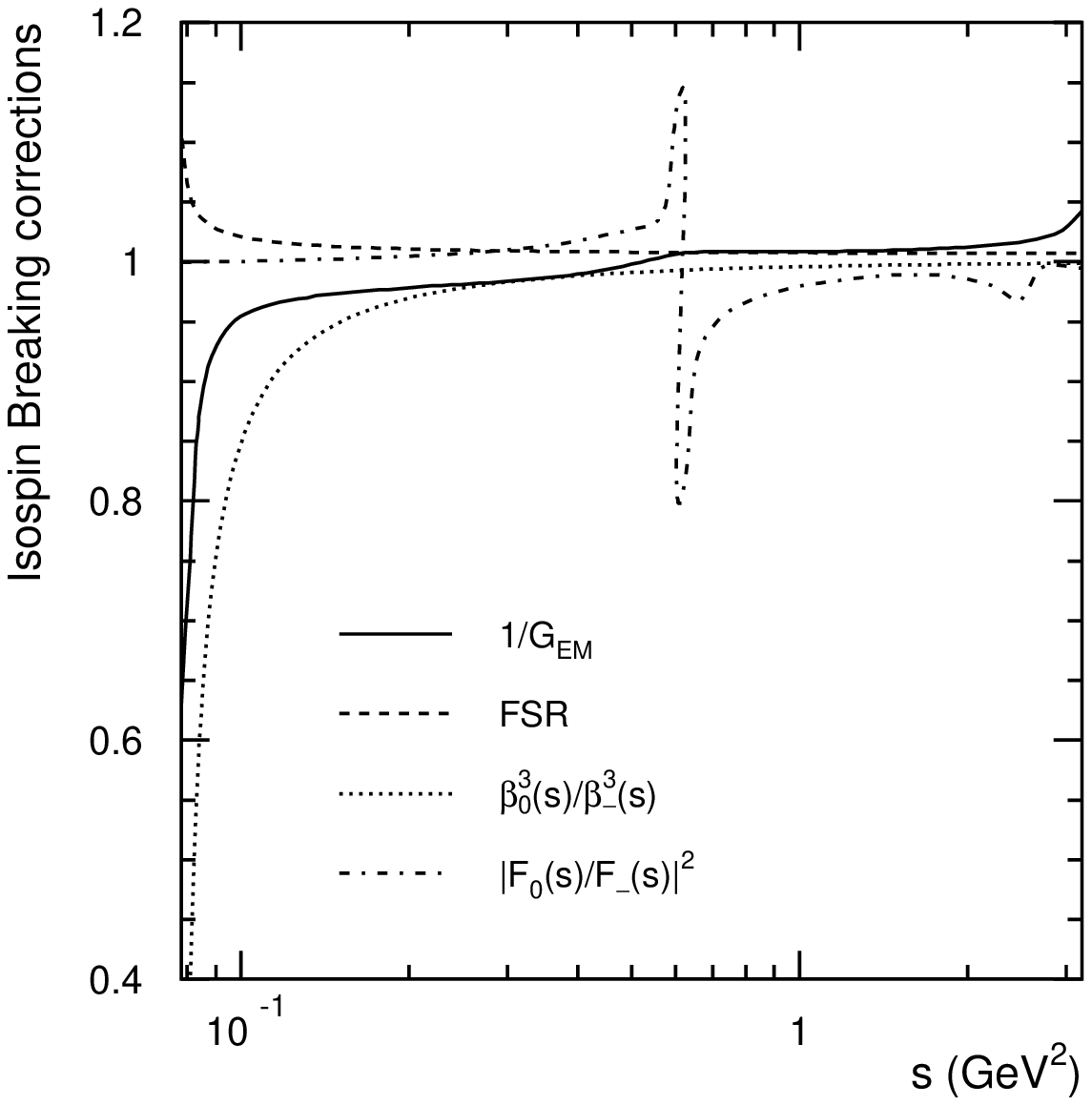} \hspace{0.3cm}
\includegraphics[width=87mm]{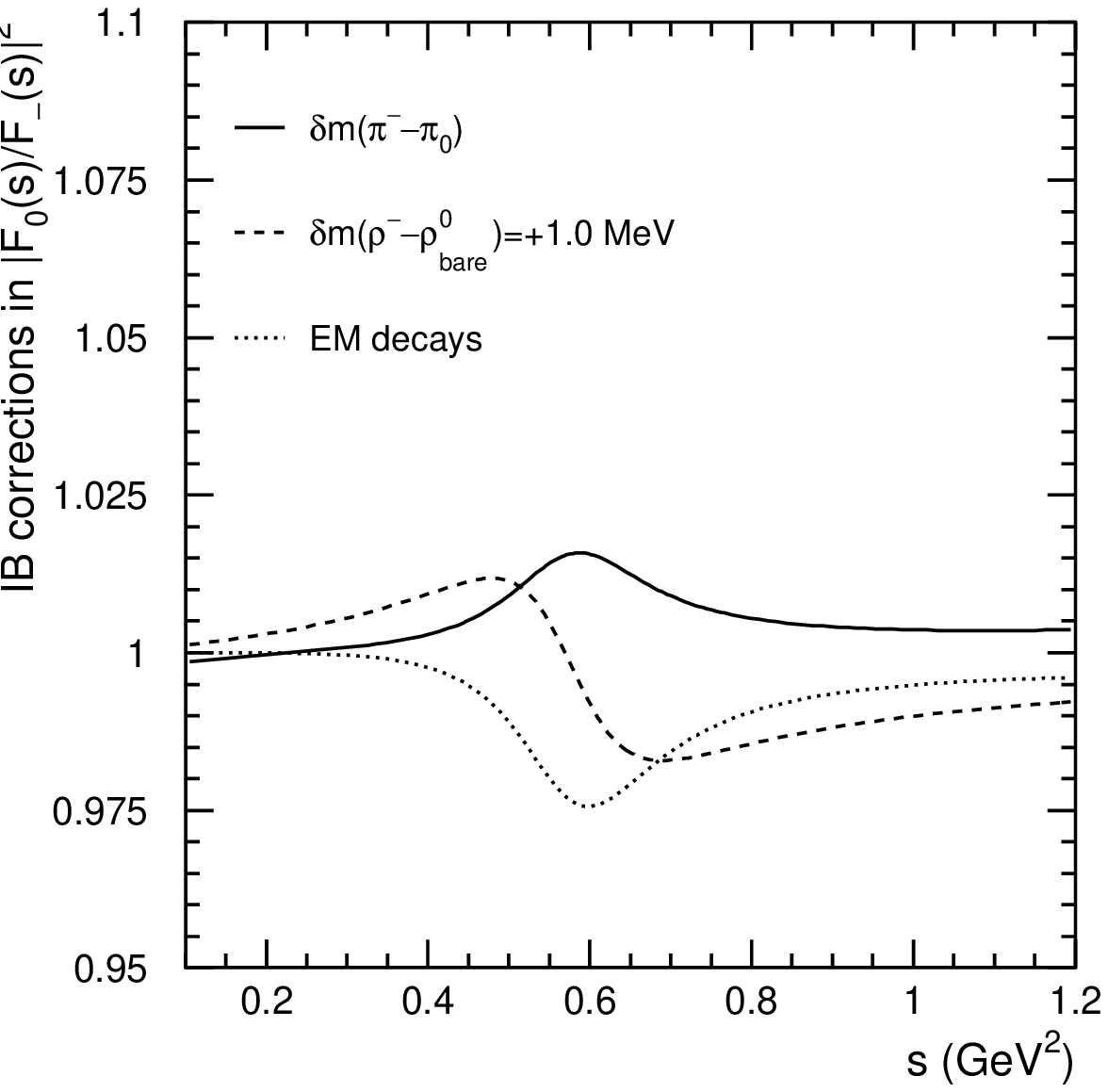}
\vspace{-0.3cm}
\caption{Left: Isospin-breaking corrections from $\Gem$, 
        FSR, $\beta^3_0(s)/\beta^3_-(s)$ and $|F_0(s)/F_-(s)|^2$. 
        Right: Isospin-breaking corrections in the ratio of $I=1$
        components of the form factors $|F_0(s)/F_-(s)|^2$ due
        to the $\pi$ mass splitting $\delta m_\pi=m_{\pi^\pm}-m_{\pi^0}$,
        the $\rho$ mass splitting 
        $\delta m_\rho=m_{\rho^\pm}-m_{\rho^0_{\rm bare}}$,
        and the difference $\delta\Gamma_\rho$ in the $\rho$ meson widths.}
\label{fig:ibcorr}
\end{figure*}
Historically, the conserved vector current (CVC) relation between \Tau and 
\ee data was considered even before the discovery of 
the \Tau lepton~\cite{tsai,sakurai}. 
In the limit of isospin invariance, the spectral function of 
the vector current decay $\tau\to X^-\nut$ is related to the $\ee\to X^0$ 
cross section of the corresponding isovector final state $X^0$,
\beq
\label{eq:cvc}
  \sigma_{X^0}^{I=1}(s) =
         \frac{4\pi\alpha^2}{s}\,v_{1,\,X^-}(s)~,
\eeq
where $s$ is the centre-of-mass energy-squared or equivalently 
the invariant mass-squared of the $\tau$ final state $X$, 
$\alpha$ is the electromagnetic fine structure constant, 
and $v_{1,\,X^-}$ is the non-strange, isospin-one vector spectral function
given by 
\beqn
\label{eq:sf}
   v_{1,\,X^-}(s)
   &=&
           \frac{m_\tau^2}{6\,|V_{ud}|^2}\,
              \frac{\BR_{X^-}}
                   {\BR_{e}}\,
              \frac{1}{N_X}\frac{d N_{X}}{ds}  \\
   & & 
              \times\,
              \left(1-\frac{s}{m_\tau^2}\right)^{\!\!-2}\!
                     \left(1+\frac{2s}{m_\tau^2}\right)^{\!\!-1}
              \frac{R_{\rm IB}(s)}{\Sew} \,, \nonumber
\eeqn
with
\begin{equation}
\label{eq:rib}
R_{\rm IB}(s)=\frac{{\rm FSR}(s)}{\Gem(s)}
              \frac{\beta^3_0(s)}{\beta^3_-(s)}
              \left|\frac{F_0(s)}{F_-(s)}\right|^2\,.
\end{equation}
In Eq.~(\ref{eq:sf}), $(1/N_X)dN_X/ds$ is the normalised invariant mass 
spectrum of the hadronic final state, and $\BR_{X^-}$ denotes 
the branching fraction of $\tau\to X^-(\gamma)\nut$ (throughout this paper, 
final state photon radiation is 
implied for $\tau$ branching fractions). We use for the $\tau$ mass the value
$m_\tau=(1776.84\pm 0.17)\,{\rm MeV}$~\cite{pdg08}, and for the CKM matrix 
element $|V_{ud}|=0.97418\pm0.00019$~\cite{ckmfitter}, 
which assumes CKM unitarity. 
For the electron branching fraction we use 
$\BR_{e}=(17.818 \pm 0.032)\%$, 
obtained~\cite{rmp} supposing lepton universality. 
Short-distance electroweak radiative effects lead to the correction 
$\Sew=1.0235\pm0.0003$~\cite{marciano,dehz02,bl90,erler04}.
All the $s$-dependent isospin-breaking (IB) corrections are included in 
$R_{\rm IB}$, and discussed in the following for the dominant $\pi\pi$ decay 
channel.

The first term in Eq.~(\ref{eq:rib}) is the ratio ${\rm FSR}(s)/\Gem(s)$, 
       where ${\rm FSR}(s)$ refers to the final state radiative corrections~\cite{fsr} in
       the $\pi^+\pi^-$ channel, and $\Gem(s)$ denotes the long-distance 
       radiative corrections of order $\alpha$ to the photon inclusive 
       $\tau^- \to \pi^-\pi^0\nu_\tau$ spectrum. 
       $\Gem(s)$ includes the virtual and real photonic corrections and was 
       calculated originally in~\cite{Cirigliano:2002pv} in the framework of 
       the Resonance Chiral Theory~\cite{portoles07cx}. In that work
       the small axial contributions to real photon emission were fixed using 
       the axial anomalous terms~\cite{Wess:1971yu}.
       A recalculation of $\Gem(s)$ was presented in~\cite{lopez}, 
       where the real photon corrections were incorporated via a meson 
       dominance model.
       Since these corrections diverge in the soft-energy limit, a small mass 
       must be given to the photon as regularisation. 
       Consistency however requires that the real photon corrections are 
       calculated by summing over all three polarisation
       states of the massive photon~\cite{Kinoshita:1958ru}.
       If we include the longitudinal polarisation according to 
       Ref.~\cite{Kinoshita:1958ru}, the model-independent piece of 
       the radiative corrections changes by at most 
       $0.3\%$ close to threshold and rapidly vanishes with increasing $s$.

The $\Gem(s)$ correction used in this analysis is based on Ref.~\cite{lopez}.
       We do not apply, however, any correction for the contribution from 
       the square of the $\pi(\omega\to\pi^0\gamma)$ amplitude, 
since it is considered
       as a background by all experiments and hence subtracted from the 
       measured spectral functions. 
       On the other hand, we do keep the interference between bremsstrahlung 
       and $\omega$ amplitudes.
       The resulting $\Gem(s)$ function is shown
       by the solid curve in the left-hand plot of Fig.~\ref{fig:ibcorr}. 
       The main numerical difference between this correction and that 
       of~\cite{Cirigliano:2002pv} lies below the $\rho$ peak. 
       Since the origin of the difference is presently only partly understood,
       we assign the full effect as systematic uncertainty to the $\Gem$
       correction.

The second correction term in Eq.~(\ref{eq:rib}), $\beta^3_0(s)/\beta^3_-(s)$,
       arises from the $\pi^\pm$--$\pi^0$ mass splitting and is important only
       close to the threshold (dotted curve in Fig.~\ref{fig:ibcorr} (left)).

The third IB correction term involves the ratio of the electromagnetic
       to weak form factors $|F_0(s)/F_-(s)|$ and is the most delicate one.
       Below $1\,{\rm GeV}$, the pion form factors are dominated by the 
       $\rho$ meson resonance, such that IB effects mainly stem from the mass 
       and width differences between the $\rho^\pm$ and $\rho^0$ mesons, 
       and from $\rho^0$--$\omega$ mixing.
       The overall effect of this correction is shown by the dash-dotted 
       curve in Fig.~\ref{fig:ibcorr} (left).

Let us analyse in more detail the IB effects in the form factors.
       A direct calculation of the $2\pi$ production amplitudes in $e^+e^-$
       annihilation and $\tau$ decays using vector meson dominance leads to
       \beqn
          F_{0}(s) &=& f_{\rho^0}(s)\left[1+\delta_{\rho \omega} 
                       \frac{s}{m_{\omega}^2-s-im_{\omega}\Gamma_{\omega}(s)} 
                                    \right]\,,\\ 
          F_{-}(s) &=& f_{\rho^-}(s)\,,
       \eeqn
       where $\delta_{\rho\omega}$ is a complex $\rho$--$\omega$ mixing 
       parameter. Following~\cite{aleph_new}, two phenomenological fits to 
       the \ee form factor data 
       have been performed using the Gounaris-Sakurai 
       (GS)~\cite{gounaris} and K\"uhn-Santamaria (KS)~\cite{ks} 
       parametrisations\footnote{
The fits are performed in the full mass range where the \ee data
are available. This differs from those fits performed in
Ref.~\cite{maltman-wolfe} in which the fits were limited to a given single 
\ee experiment (with data available only below $1\,{\rm GeV}$) 
with fewer number of free parameters. 
We do not use the Hidden Local Symmetry effective
model~\cite{benayoun} and the effective field theory model~\cite{pich}
as these models do not include contributions from the high mass resonances
such as $\rho^\prime$ and therefore can only be valid for the mass range below
about $1\,{\rm GeV}$.
}.
       For the corresponding 
       mixing strengths and phases of the fits we find 
       $|\delta^{\rm GS}_{\rho\omega}|=(2.00\pm 0.06)\times 10^{-3}$, 
       ${\rm arg}(\delta^{\rm GS}_{\rho\omega})=(11.6\pm 1.8)^\circ$, and
       $|\delta^{\rm KS}_{\rho\omega}|=(1.87\pm 0.06)\times 10^{-3}$, 
       ${\rm arg}(\delta^{\rm KS}_{\rho\omega})=(13.2\pm 1.7)^\circ$, 
       respectively.
       In both parametrisations, the absorptive parts of the $\rho$ propagators
       have an explicit energy-dependence of the form 
       $-i\sqrt{s}\Gamma_{\rho^{0,-}}(s)$.

One of the IB effects is associated with the $\rho$ meson width difference. 
       Within an accuracy of $0.1\%$, the decay widths of the $\rho$ mesons 
       below $\sqrt{s}=1\,{\rm GeV}$ are given by their photon inclusive rates
       into $\pi\pi$ modes~\cite{lopez-width}. A direct calculation of the 
       $\rho\to \pi\pi(\gamma)$ and $\pi\pi\gamma$ decay rates shows that the
       width difference, $\delta\Gamma_\rho = \Gamma_{\rho^0}-\Gamma_{\rho^-}$,
       is given by~\cite{lopez-width}
       \beq
       \label{wdiff} 
          \delta\Gamma_\rho(s) = \frac{g_{\rho\pi\pi}^2\sqrt{s}}{48\pi} 
             \left[\beta_0^3(s)(1+\delta_0)-\beta_-^3(s)(1+\delta_-) \right]\,,
       \eeq
       where $g_{\rho\pi\pi}$ is the strong coupling of the isospin-invariant
       $\rho\pi\pi$ vertex and $\delta_{0,-}$ denote radiative corrections
       for photon-inclusive $\rho\to \pi\pi$ decays, which include 
       $\rho\to \pi\pi\gamma$. Contrary to expressions used in previous 
       approaches the $\rho$ meson decay widths in Eq.~(\ref{wdiff}) are 
       independent of the photon energy cut-off used to separate the 
       $\rho\to\pi\pi(\gamma)$ and $\rho\to\pi\pi\gamma$ rates.
       In addition to the IB arising from the $\pi^\pm$--$\pi^0$ mass 
       difference, the radiative corrections to $\rho\to\pi\pi$ and their 
       corresponding radiative rates produce a splitting in the $\rho$ meson 
       widths.
       For instance, at $\sqrt{s}=m_\rho=775\,{\rm MeV}$, the width difference
       of Eq.~(\ref{wdiff}) is $\delta\Gamma_{\rho}\approx +0.76\,{\rm MeV}$, 
       compared to the value 
       $\delta \Gamma_\rho\approx (-0.42\pm 0.58)\,{\rm MeV}$ used 
       in~\cite{adh}.
       The difference between the two results is mainly due to the effects 
       from radiative corrections (the $\delta_{0,-}$ terms in 
       Eq.~(\ref{wdiff})). Our results can also be compared to the one used 
       in~\cite{Cirigliano:2002pv}, $\delta \Gamma_\rho=(m_\rho s/96\pi F^2_\pi)[\beta^3_0(s)-\beta^3_-(s)]+(0.45\pm 0.45)\,{\rm MeV}$,
       which at $\sqrt{s}=775\,{\rm MeV}$ gives 
       $\delta\Gamma_\rho=(-0.61\pm 0.45)\,{\rm MeV}$. 
       Note that if electromagnetic effects were ignored ($\delta_{0,-}=0$)
       in Eq.~(\ref{wdiff}), we would have $\delta \Gamma=-1.06\,{\rm MeV}$,
       which is very similar to the cases considered 
       in~\cite{Cirigliano:2002pv,adh}.

The second input required to assess the IB effects in the form factors is
       the mass splitting between neutral and charged $\rho$ mesons. 
       Using the expected difference 
       $m_{\rho^0}-m_{\rho^0_{\rm bare}}\approx 3\Gamma(\rho^0\to \ee)/(2\alpha)=1.45\,{\rm MeV}$,
       between dressed and bare $\rho^0$ mass~\cite{sakurai69}, 
       together with the experimental 
       value $m_{\rho^\pm}-m_{\rho^0}=(-0.4\pm 0.9)\,{\rm MeV}$, 
       obtained by KLOE from a fit to the $\phi\to\pi^+\pi^-\pi^0$ Dalitz 
       plot~\cite{kloe-phi-rhopi}, one finds 
       $\delta m_\rho=m_{\rho^\pm}-m_{\rho^0_{\rm bare}}=(1.0\pm 0.9)\,{\rm MeV}$,
       which we use here instead of the degeneracy assumed in previous 
       analyses~\cite{Cirigliano:2002pv,dehz02}.

The IB effects in the ratio of $I=1$ components of the pion form factors 
       (except for $\rho$--$\omega$ mixing) are drawn in the right-hand plot 
       of Fig.~\ref{fig:ibcorr}. 
       It is noticeable that the effects of photonic corrections and of the 
       $\pi^\pm$--$\pi^0$ mass difference in the $\rho$ meson widths largely 
       cancel each other.

\begin{figure*}[t]
\includegraphics[width=89mm]{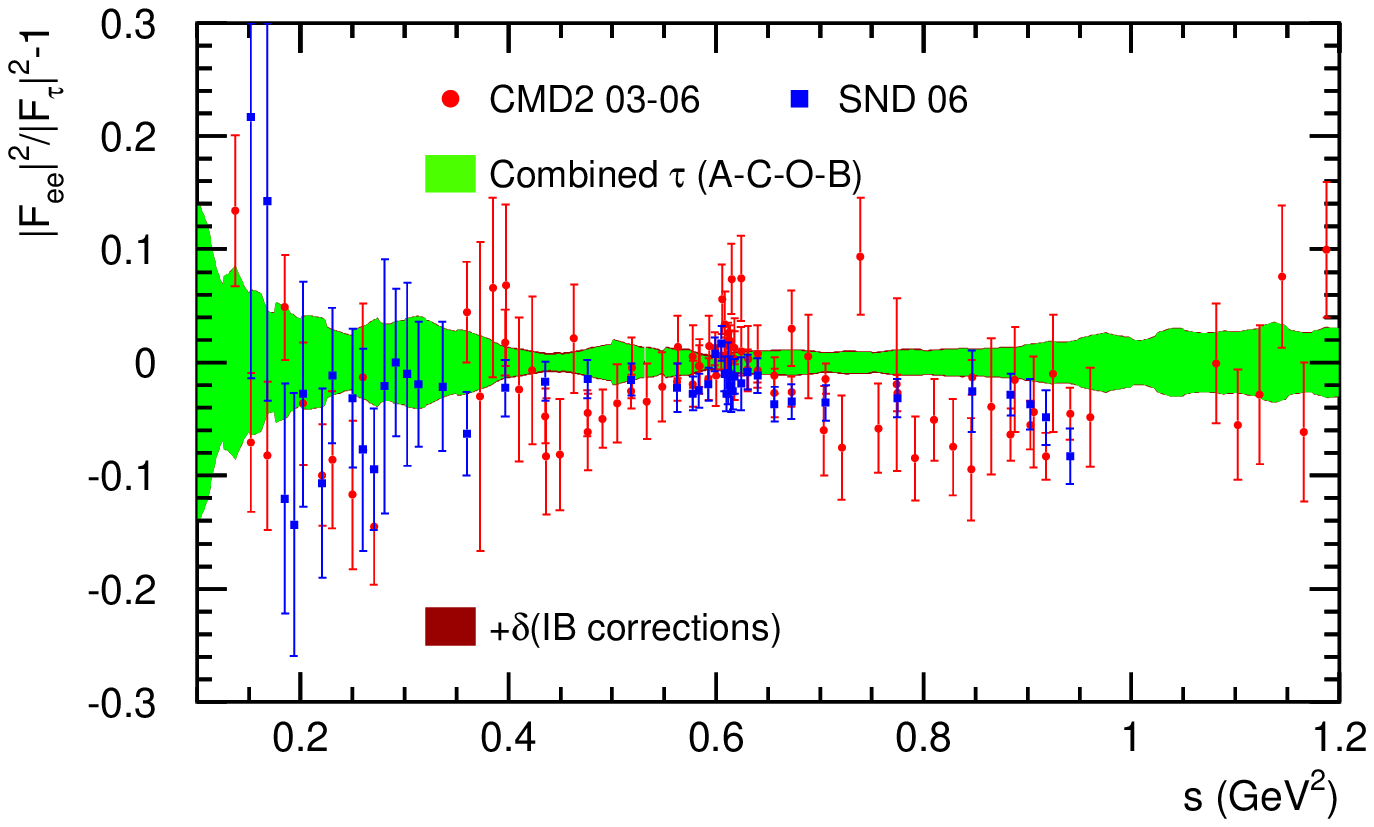}
\includegraphics[width=89mm]{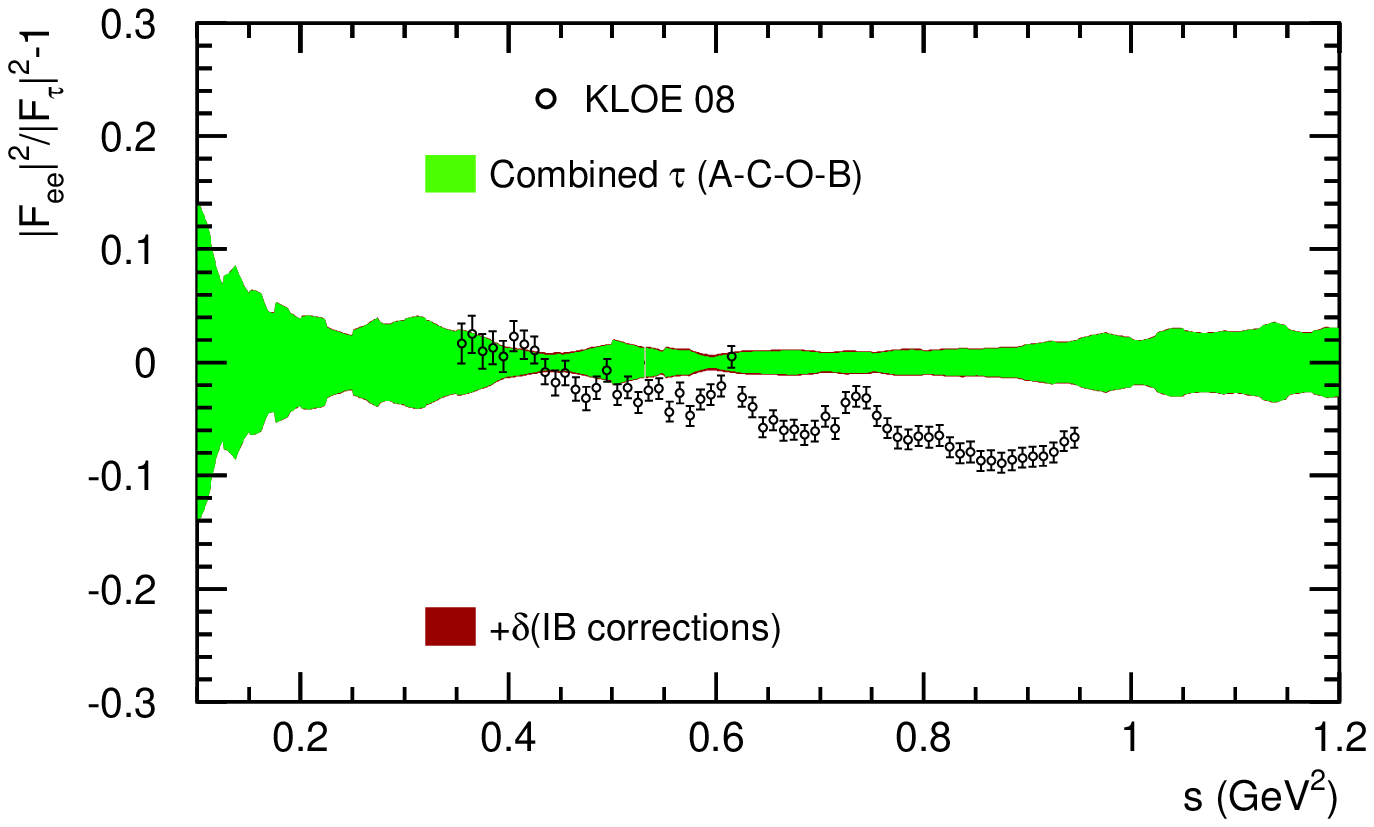}
\vspace{-0.4cm}
\caption{Relative comparison between \ee and \Tau spectral functions, 
   expressed in terms of the difference between neutral and charged pion form 
   factors. Isospin-breaking (IB) corrections are applied to \Tau data with 
   its uncertainties, although hardly visible, included in the error band.}
\label{fig:tauee}
\end{figure*}

Figure~\ref{fig:tauee} shows the relative difference between the \ee 
and the isospin-breaking-corrected \Tau spectral functions versus $s$. 
The relative normalisation is consistent within the respective errors
and the shape is found in better agreement than before~\cite{dehz03}, despite 
a remaining deviation above the $\rho$-mass-squared. The discrepancy with 
the KLOE data, although reduced, persists.

\section{Update of $a_\mu^{\rm had,LO}[\pi\pi, \tau]$}

\begin{table}[htb]
  \caption[.]{\label{tab:amu}
    Contributions to $\amuhadLO~[\pi\pi, \tau]$ ($\times10^{-10}$)
    from the isospin-breaking corrections discussed in Sec.~\ref{sec:ib}.
    Corrections shown in two separate columns correspond to the 
    Gounaris-Sakurai (GS) and K\"uhn-Santamaria (KS) parametrisations,
    respectively.} 
\setlength{\tabcolsep}{0.0pc}
\begin{tabularx}{\columnwidth}{@{\extracolsep{\fill}}lcc} 
\hline\noalign{\smallskip}
      & \multicolumn{2}{c}{$\Delta \amuhadLO[\pi\pi, \tau]$ ($10^{-10}$)} \\
 \rs{Source}  & GS model & KS model \\
\noalign{\smallskip}\hline\noalign{\smallskip}
\Sew                &  \multicolumn{2}{c}{$-12.21\pm0.15$}  \\
$\Gem$              &  \multicolumn{2}{c}{$ -1.92\pm0.90$}  \\
FSR                 &  \multicolumn{2}{c}{$+4.67\pm0.47$}   \\
$\rho$--$\omega$ interference
                    & $+2.80\pm 0.19$ & $+2.80\pm 0.15$  \\
$m_{\pi^\pm}-m_{\pi^0}$ effect on $\sigma$
                    &  \multicolumn{2}{c}{$ -7.88$}        \\
$m_{\pi^\pm}-m_{\pi^0}$ effect on $\Gamma_{\rho}$
                    & $+4.09$ & $ +4.02$   \\
$m_{\rho^\pm}-m_{\rho^0_{\rm bare}}$ 
                    & $0.20^{+0.27}_{-0.19}$ & $ 0.11^{+0.19}_{-0.11}$    \\
$\pi\pi\gamma$, electrom. decays
                    & $ -5.91\pm0.59$ & $-6.39\pm 0.64$ \\
\noalign{\smallskip}\hline\noalign{\smallskip}
               & $-16.07\pm 1.22$ & $-16.70\pm 1.23$ \\
\rs{Total}               & \multicolumn{2}{c}{$-16.07\pm 1.85$}\\
\noalign{\smallskip}\hline
\end{tabularx}
\end{table}
\begin{table*}[htb]
  \caption[.]{\label{tab:amu_tau}
    The IB-corrected $\amuhadLO~[\pi\pi, \tau]$ ($\times10^{-10}$) from the 
    measured mass spectrum by ALEPH, CLEO, OPAL and Belle, and the combined 
    spectrum using the corresponding branching fraction values. The results 
    are shown separately in two different energy ranges. 
    The first errors are due 
    to the shapes of the mass spectra, which also include a small contribution 
    of $0.11$ from the $\tau$ mass and $|V_{ud}|$. 
    The second errors are due to 
    $\BR_{\pi\pi^0}$ and $\BR_e$, and the third errors from the 
    isospin-breaking corrections, which are partially anti-correlated between
    the two energy ranges. The last line gives the world average 
    branching fraction and also the evaluations of the combined spectra 
    (which are not equivalent to the arithmetic averages of the individual 
    evaluations -- see text).}
\setlength{\tabcolsep}{0.0pc}
\begin{tabularx}{\textwidth}{@{\extracolsep{\fill}}cccc} 
\hline\noalign{\smallskip}
Experiment &  \multicolumn{2}{c}{$\amuhadLO[\pi\pi, \tau]$ ($10^{-10}$)} & 
  $\BR_{\pi\pi^0}$ (\%) \\
           &  $2m_{\pi^\pm}-0.36\,{\rm GeV}$ & $0.36-1.8\,{\rm GeV}$ & \\
\noalign{\smallskip}\hline\noalign{\smallskip}
ALEPH &  $9.46\pm 0.33_{\rm exp} \pm 0.05_\BR\pm 0.07_{\rm IB}$ 
      & $499.19\pm 5.20_{\rm exp}\pm 2.70_\BR\pm 1.87_{\rm IB}$ &
  $25.49\pm 0.10_{\rm stat}\pm 0.09_{\rm syst}$ \\
CLEO  &  $9.65\pm 0.42_{\rm exp}\pm 0.17_\BR\pm 0.07_{\rm IB}$
      & $504.51\pm 5.36_{\rm exp}\pm 8.77_\BR\pm 1.87_{\rm IB}$ &
  $25.44\pm 0.12_{\rm stat}\pm 0.42_{\rm syst}$ \\
OPAL  &  $11.31\pm 0.76_{\rm exp} \pm 0.15_\BR\pm 0.07_{\rm IB}$
      & $515.56\pm 9.98_{\rm exp}\pm 6.95_\BR\pm 1.87_{\rm IB}$ & 
  $25.46\pm 0.17_{\rm stat}\pm 0.29_{\rm syst}$ \\
Belle &  $9.74\pm 0.28_{\rm exp}\pm 0.15_\BR\pm 0.07_{\rm IB}$ 
      & $503.95\pm 1.90_{\rm exp}\pm 7.84_\BR\pm 1.87_{\rm IB}$ &
  $25.24\pm 0.01_{\rm stat}\pm 0.39_{\rm syst}$ \\
\noalign{\smallskip}\hline\noalign{\smallskip}
Combined &  $9.76\pm 0.14_{\rm exp} \pm 0.04_\BR\pm 0.07_{\rm IB}$ 
         & $505.46\pm 1.97_{\rm exp}\pm 2.19_\BR\pm 1.87_{\rm IB}$ & 
  $25.42\pm 0.10$ \\
\noalign{\smallskip}\hline
\end{tabularx}
\end{table*}
The IB corrections applied to the lowest order hadronic contribution
to the muon $g-2$ using $\tau$ data in the dominant
$\pi\pi$ channel can be evaluated with
\beqns
\lefteqn{\Delta^{\rm IB} a^{\rm LO, had}_\mu[\pi\pi, \tau]=
           \frac{\alpha^2 m_\tau^2}{6\,|V_{ud}|^2 \pi^2}\,
           \frac{\BR_{\pi\pi^0}}{\BR_{e}}\,
           \int^{m^2_\tau}_{4m^2_\pi} ds \,
              \frac{K(s)}{s}} \nonumber\\
            &&\times\,
              \frac{d N_{X}}{N_X\,ds}\,
              \left(1-\frac{s}{m_\tau^2}\right)^{\!\!-2}\!
                     \left(1+\frac{2s}{m_\tau^2}\right)^{\!\!-1} 
              \left[\frac{R_{\rm IB}(s)}{\Sew}-1\right]\,,
\eeqns
where $K(s)$ is a QED kernel function~\cite{br68}.

The numerical values for the various corrections are given in 
Table~\ref{tab:amu} for the energy range between the $2\pi$ mass threshold
and $1.8\,{\rm GeV}$.
The present estimate of the IB effect from long-distance corrections
is smaller than the previous one~\cite{md_tau06,lopez}, because
we now use a $\Gem(s)$ correction in which the contributions involving 
the $\rho\omega\pi$ vertex are explicitly excluded (except for its 
interference with the QED amplitude).
Its uncertainty corresponds to the difference between the correction used 
in this analysis and that from Ref.~\cite{Cirigliano:2002pv}.
The quoted $10\%$ uncertainty on the FSR and $\pi\pi\gamma$
electromagnetic corrections is
an estimate of the structure-dependent effects (pion form factor) in virtual 
corrections and of intermediate resonance contributions to real
photon emission~\cite{lopez-width, fl08, dub05}.
The systematic uncertainty assigned to the $\rho$--$\omega$ interference 
contribution accounts for the difference in \amuhadLO between 
two phenomenological fits, where the mass and width of the $\omega$ resonance 
are either left free to vary or fixed to their world average values.

Some of the corrections in Table~\ref{tab:amu} are parametrisation dependent.
We choose to take the final corrections from the Gounaris-Sakurai parametrisation and assign the full difference with respect to the KS results\footnote
{
We do not confirm the significant IB correction difference of the KS 
parametrisation on the $\rho-\omega$ interference with respect to the GS 
parametrisation observed in Ref.~\cite{maltman-wolfe}.
} 
as systematic error. The total correction for isospin breaking amounts to
$(-16.07\pm 1.85)\cdot10^{-10}$ for $\amuhadLO[\pi\pi,\tau]$, 
where all systematic errors have been added in quadrature except for
the GS and KS difference which has been added linearly. This correction 
is to be compared to the value $(-13.8\pm2.4)\cdot10^{-10}$ obtained 
previously~\cite{dehz02}.
Since the FSR correction was previously included, but not counted in the IB 
corrections, the net change amounts to
$-6.9\times 10^{-10}$, dominated by the electromagnetic decay correction.

\begin{figure*}[htb]
\includegraphics[width=89mm]{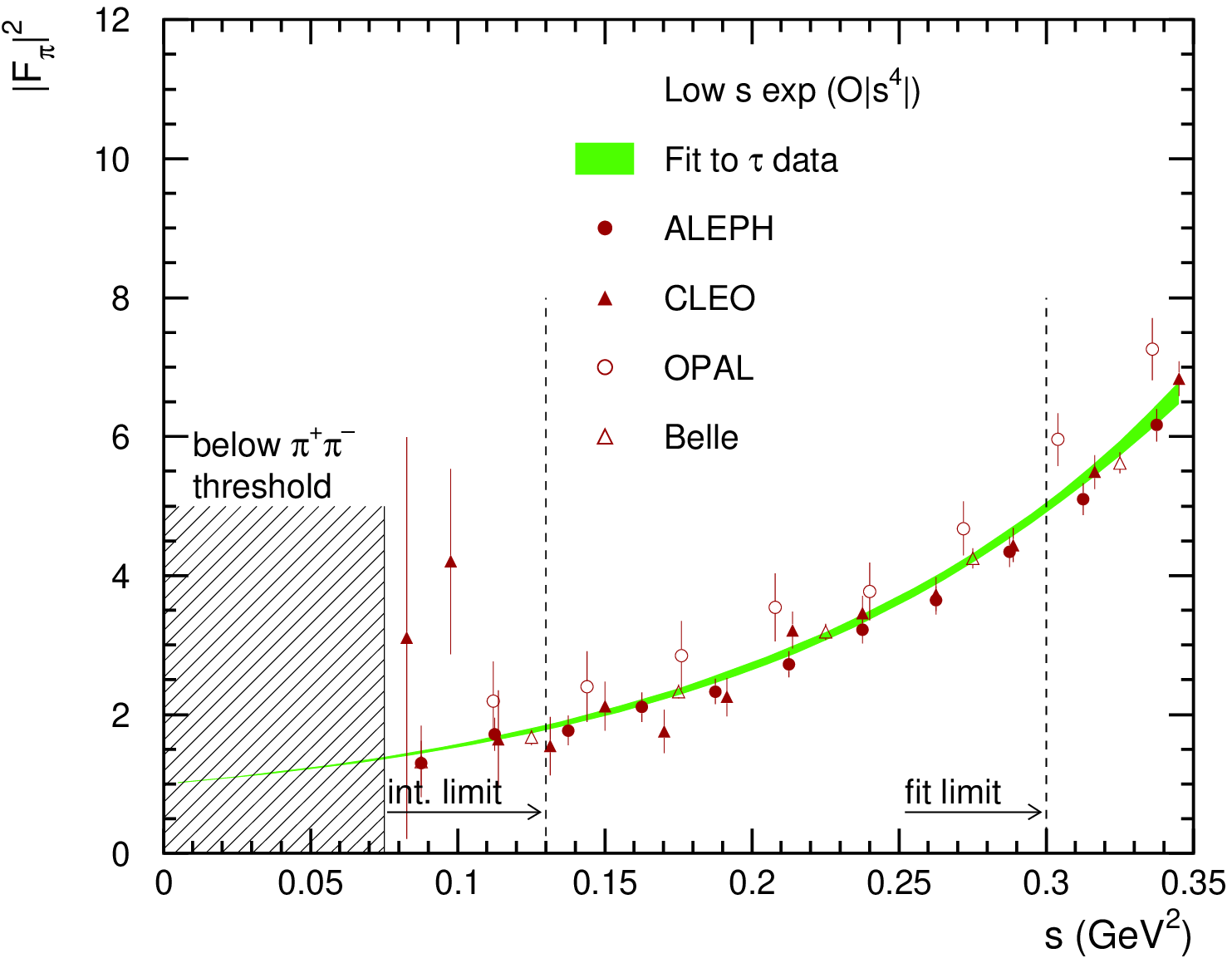} 
\includegraphics[width=89mm]{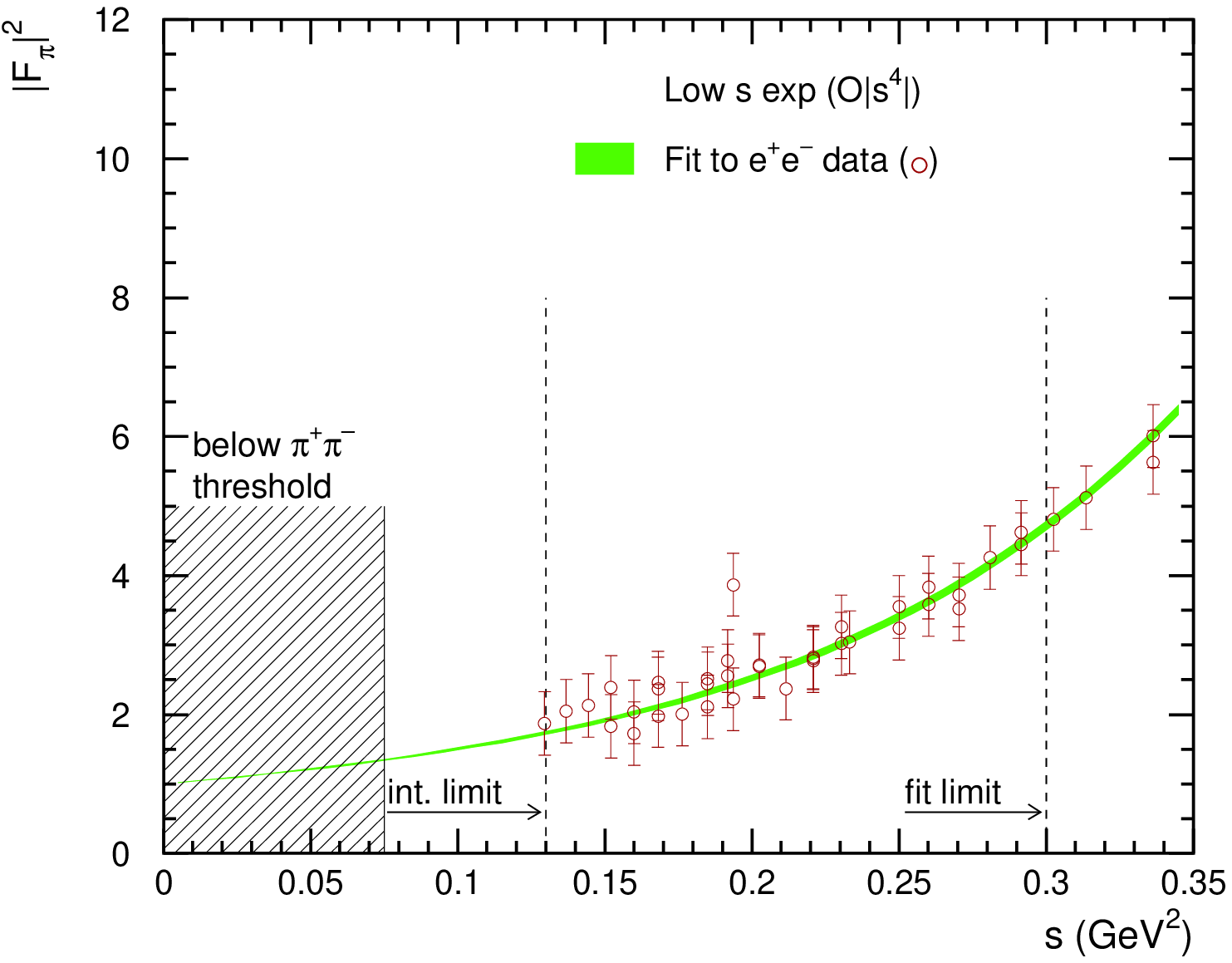}
\vspace{-0.3cm}
\caption{Fit of the pion form factor from $4m^2_\pi$ to $0.3\,{\rm GeV}^2$
         using a third order expansion with the constraints $F(0)=1$ and using 
         the measured
         pion charge radius-squared from space-like data~\cite{amendolia86}. 
         The result of the fit to the $\tau$ data (left) and to $e^+e^-$
         data (right) is integrated only up to $0.13\,{\rm GeV}^2$, beyond which we
         directly integrate over the data points. }
\label{fig:lowen_fit}
\end{figure*}
The corresponding IB-corrected $a^{\rm had, LO}_\mu[\pi\pi, \tau]$ 
in the dominant 
$\pi^+\pi^-$ channel below 1.8\gev is given in Table~\ref{tab:amu_tau} 
for ALEPH, CLEO, OPAL, Belle, and for the combined mass 
spectrum from these experiments. 
The evaluation at energy below $0.36\,{\rm GeV}$ is obtained
by fitting an expansion in $s$ to the corresponding mass spectrum 
following the method introduced in~\cite{dehz02}.
The comparison of the fit with the $\tau$ data at low energy
is shown in Fig.~\ref{fig:lowen_fit} (left). Good agreement is observed. 
Indeed, a direct determination using data gives 
$(10.18\pm 0.98_{\rm exp})\times 10^{-10}$ in agreement with
the fit-based result of $(9.76\pm 0.14_{\rm exp})\times 10^{-10}$, 
which is more precise because of the constraint $F(0)=1$.
The evaluation in the remaining energy region is performed directly
from a finely ($1\,{\rm MeV}$) binned mass spectrum obtained using HVPTools 
by interpolating the original measurements with second order polynomials 
(conserving by means of renormalisation the integral in each bin before and 
after interpolation).
The consistent propagation of all errors is ensured by generating
large samples of pseudo Monte Carlo experiments.
The uncertainty due to the interpolation procedure is estimated from 
a test with a known model to be at most $0.2\times 10^{-10}$, 
which is negligible compared to the other systematic uncertainties.
It is interesting to compare the first and second 
errors between the experiments. The first errors are mainly experimental, but also 
include small contributions from the uncertainties in $\tau$ mass and 
$|V_{ud}|$.
The second errors are due to $\BR_{\pi\pi^0}$, as measured by each 
experiment, and -- to a lesser extent -- to $\BR_e$, for which a common
value has been used everywhere. Belle has the most precise experimental 
precision on the measurement of the mass spectrum,
whereas ALEPH dominates the $\BR_{\pi\pi^0}$ measurement. The result 
$a^{\rm had, LO}_\mu[\pi\pi,\tau]=515.2\pm 2.0_{\rm exp}\pm 2.2_\BR\pm 1.9_{\rm IB}$ 
(if not stated otherwise, this and the following numbers for $a_\mu$ 
are given in units of $10^{-10}$)
is obtained from the combined $\pi^-\pi^0$ mass spectrum of ALEPH, CLEO, 
OPAL and Belle using the world average 
$\BR_{\pi\pi^0}=(25.42\pm 0.10)\%$.
This result is consistent with
the direct average
$516.1\pm 1.8_{\rm exp}\pm 2.2_\BR\pm 1.9_{\rm IB}$, 
obtained from the four individual $a_\mu$ calculations.
The experimental error from the combined spectrum is slightly less 
precise as it accounts for the incompatibility between experiments 
in certain region of the mass spectrum. 

The contributions to \amuhadLO from the $\pi^+\pi^-2\pi^0$ and $2\pi^+2\pi^-$ 
channels below 1.8\gev are $21.4\pm 1.3_{\rm exp}\pm 0.6_{\rm IB}$ and 
$12.3\pm1.0_{\rm exp}\pm0.4_{\rm IB}$, respectively. This leads to the 
complete \Tau-based lowest order hadronic contribution 
\beqn
   \amuhadLO[\Tau] &=& 705.3 \pm 3.9_{\rm exp}\pm 0.7_{\rm rad} \pm 0.7_{\rm QCD}\pm 2.1_{\rm IB}\,,
                       \nonumber\\
                   &=& 705.3 \pm 4.5\,, 
\eeqn
where the second error is due to our treatment of (potentially) missing 
radiative corrections in old data included in the calculation of 
the dispersion integral~\cite{dehz03}, 
and the third error stems from the uncertainty in the perturbative evaluation 
of the inclusive hadronic cross section in the energy ranges 
$1.8$--$3.7\,{\rm GeV}$ and beyond $5\,{\rm GeV}$. 
The central value decreases from previously $710.3$, 
obtained using incomplete isospin corrections~\cite{md_tau06} 
and the superseded combined $\tau$ spectral function from ALEPH, CLEO and OPAL.

\begin{table*}[htb]
  \caption[.]{\label{tab:amu_ee}
    Evaluated $\amuhadLO~[\pi\pi, \ee]$ ($\times10^{-10}$) contribution 
    from the $\ee$ data, including and excluding KLOE data. The errors
    correspond to the experimental uncertainties with the statistical 
    and systematical errors added in quadrature (but shown separately for 
    individual experiments).}
\setlength{\tabcolsep}{0.0pc}
\begin{tabularx}{\textwidth}{@{\extracolsep{\fill}}cccc} 
\hline\noalign{\smallskip}
                          &                 &  \mc{2}{c}{$\amuhadLO[\pi\pi, \ee]$ ($10^{-10}$)} \\
  \rs{Energy range (GeV)} & \rs{Experiment} &  Incl. KLOE & Excl. KLOE \\
\noalign{\smallskip}\hline\noalign{\smallskip}
$2m_{\pi^\pm}-0.36$ & Combined $\ee$ (fit)  & \mc{2}{c}{$9.71\pm 0.12_{\rm exp}$} \\ 
\noalign{\smallskip}\hline\noalign{\smallskip}
$0.36-0.63$   & Combined $\ee$ & $120.27\pm 1.67_{\rm exp}$ & $119.63\pm 1.88_{\rm exp}$ \\ 
\noalign{\smallskip}\hline\noalign{\smallskip}
$0.63-0.958$ & CMD2 03   & \mc{2}{c}{$361.82\pm 2.43_{\rm stat}\pm 2.10_{\rm syst}$}\\
             & CMD2 06   & \mc{2}{c}{$360.17\pm 1.75_{\rm stat}\pm 2.83_{\rm syst}$}\\
             & SND  06   & \mc{2}{c}{$360.68\pm 1.38_{\rm stat}\pm 4.67_{\rm syst}$}\\
             & KLOE 08   & \mc{2}{c}{$356.82\pm 0.39_{\rm stat}\pm 3.08_{\rm syst}$}\\
             & Combined $\ee$ & $358.51\pm 2.41_{\rm exp}$ & $360.24\pm 3.02_{\rm exp}$ \\ 
\noalign{\smallskip}\hline\noalign{\smallskip}
$0.958-1.8$  & Combined $\ee$ & $15.02\pm 0.36_{\rm exp}$  & $15.02\pm 0.39_{\rm exp}$ \\ 
\noalign{\smallskip}\hline\noalign{\smallskip}
Total        & Combined $\ee$ & $503.51\pm 3.47_{\rm exp}$ & $504.60\pm 4.33_{\rm exp}$ \\
\noalign{\smallskip}\hline
\end{tabularx}
\end{table*}
We also re-evaluate the lowest order hadronic contribution to the muon $g-2$ 
using \ee data, updating our most recent preliminary result~\cite{md_tau06} 
with published CMD-2~\cite{cmd2new} and KLOE~\cite{kloe08} data. 
The results are given in Table~\ref{tab:amu_ee}. 
We have separated the evaluation into four distinct energy ranges. 
The most recent $\ee$ data from CMD2, SND and KLOE overlap in the range 
$0.63$--$0.958\,{\rm GeV}$ so that the corresponding 
$a_\mu^{\rm had, LO}[\pi\pi,\ee]$ values can be compared. 
Agreement is observed between CMD2 and SND, while KLOE lies somewhat lower. 
To account for this, we consider two combinations of the $\ee$ data, 
distinguished by either including or excluding the KLOE data. 
The combination of the data is performed also using HVPTools~\cite{hvptools},
to transform the original $\ee$ bare cross sections and 
associated statistical and systematic covariance matrices into 
fine-grained energy bins ($1\,{\rm MeV}$), taking 
into account to our best knowledge the correlation within each experiment 
as well as between the experiments. 
The evaluation in the low energy range 
$2m_{\pi^\pm}$--$0.36\,{\rm GeV}$ is performed as for the $\tau$ data 
by fitting an expansion 
in $s$ to the combined $\ee$ data~\cite{dehz02} (right-hand plot of 
Fig.~\ref{fig:lowen_fit}), benefiting from additional space-like precision 
data~\cite{amendolia86}. The evaluations in the other three
energy ranges are obtained by 
integrating directly the combined $\ee$ cross sections
(\cf\ Table~\ref{tab:amu_ee}).

We find for the difference, $\delta a_\mu^{\rm had,LO}[\pi\pi]$, 
between the \Tau and \ee-based evaluations in the dominant $\pip\pim$ channel
\beq
   \delta a_\mu^{\rm had,LO}[\pi\pi]=
       \left\{
       \begin{array}{ll}
            11.7 \pm 3.5_{ee} \pm 3.5_{\tau+{\rm IB}}\,,\\
            10.6 \pm 4.3_{ee} \pm 3.5_{\tau+{\rm IB}}\,,
       \end{array}
   \right.
\eeq
where the upper (lower) value is for KLOE data included (excluded). 
The discrepancies amount to $2.4$ and $1.9$ times the overall errors, 
respectively.

Including the contributions from the other hadronic channels~\cite{md_tau06}, 
we find for the total \ee-based lowest order hadronic evaluation 
\beqns
   a_\mu^{\rm had, LO}[\ee]=
       \left\{
       \begin{array}{ll}
            689.8\pm4.3_{{\rm exp}+{\rm rad}}\pm 0.7_{\rm QCD}\,,\\
            690.9\pm5.2_{{\rm exp}+{\rm rad}}\pm 0.7_{\rm QCD}\,, &
       \end{array}
\right.
\eeqns
with total errors of $4.4 (5.2)$ when including (excluding) KLOE. 
Adding the other contributions~\cite{md_tau06} including the latest
estimate of the light-by-light scattering (LBLS) contribution of
$10.5\pm 2.6$~\cite{prades09}, 
we obtain the Standard Model predictions (still in $10^{-10}$ units)
\beqns
  \amuSM[\tau]           &=& 11\,659\,193.2 \pm 4.5 \pm 2.6 \pm 0.2\;,\\
  \amuSM[\ee]            &=& 
     \left\{
     \begin{array}{ll}
         11\,659\,177.7 \pm 4.4 \pm 2.6 \pm 0.2\,, \\
         11\,659\,178.8 \pm 5.2 \pm 2.6 \pm 0.2\,, 
     \end{array}
  \right.
\eeqns
where the first errors are due to the lowest order hadronic contributions, 
the second error includes higher hadronic orders, 
dominated by the uncertainty in the LBLS contribution, 
and the third error accounts for 
the uncertainties in the electromagnetic and weak contributions. 
The predictions deviate from the experimental average, 
$\amuExp=11\,659\,208.9(5.4)(3.3)$~\cite{bnl, pdgg-2rev}, 
by $15.7 \pm 8.2$ ($\tau$), 
$31.2\pm 8.1$ (\ee with KLOE) and $30.1 \pm 8.6$ (\ee without KLOE), 
respectively. 

\begin{figure}[htb]
\includegraphics[width=\columnwidth]{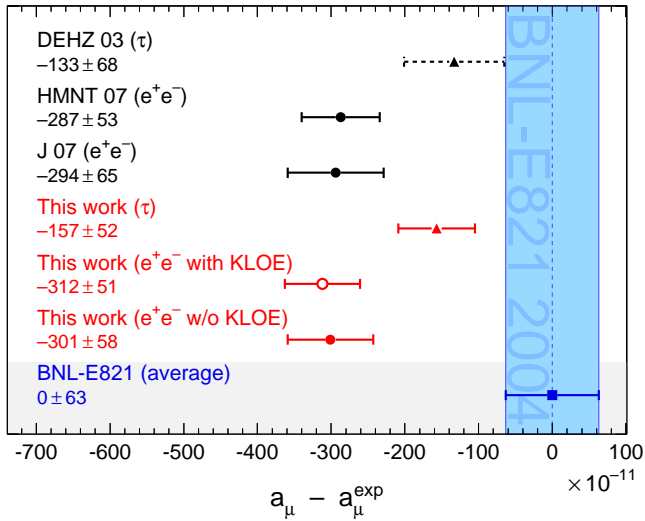}
\vspace{-0.3cm}
\caption{Compilation of recently published results for $\amuSM$ 
        (in units of $10^{-11}$),
        subtracted by the central value of the experimental 
        average~\cite{bnl, pdgg-2rev}.
        The shaded band indicates the experimental error. 
        The SM predictions are taken from: 
        DEHZ 03~\cite{dehz03}, HMNT 07~\cite{hmnt}, 
        J 07~\cite{jeger},
        and the present \Tau- and \ee-based predictions using \Tau and \ee
        spectral functions. }
\label{fig:amures}
\end{figure}
The lowest order hadronic contribution now reaches an uncertainty 
that is smaller than the measurement error and comparable in size with
the LBLS uncertainty. 
Further progress in this field thus requires, 
apart from continuously improved low-energy \ee cross section measurements, 
a more accurate muon $g-2$ measurement and LBLS calculation. 
A compilation of this and other recent $\amuSM$ predictions, compared to 
the experimental value, is shown in Fig.~\ref{fig:amures}.

\section{CVC prediction of $\BR_{\pi\pi^0}$}
\begin{table}[htb]
  \caption[.]{\label{tab:ib_br}
    Contributions to \BCVCppz ($\times10^{-2}$) 
    from the isospin-breaking corrections discussed in Sec.~\ref{sec:ib}.
    For those corrections shown in two separated columns, they correspond
    to the Gounaris-Sakurai and K\"uhn-Santamaria 
    parametrisations, respectively.}
\setlength{\tabcolsep}{0.0pc}
\begin{tabularx}{\columnwidth}{@{\extracolsep{\fill}}lcc} 
\hline\noalign{\smallskip}
  & \multicolumn{2}{c}{$\Delta \BCVCppz$ ($10^{-2}$)} \\
 \rs{Source}  &  GS model & KS model \\
\noalign{\smallskip}\hline\noalign{\smallskip}
\Sew                & \multicolumn{2}{c}{$+0.57\pm0.01$} \\
$\Gem$              & \multicolumn{2}{c}{$-0.07\pm0.17$} \\
FSR                 & \multicolumn{2}{c}{$-0.19\pm0.02$} \\
$\rho$--$\omega$ interference
                        & $-0.01\pm 0.01$ & $-0.02\pm 0.01$ \\
$m_{\pi^\pm}-m_{\pi^0}$ effect on $\sigma$
                        & \multicolumn{2}{c}{$+0.19$} \\
$m_{\pi^\pm}-m_{\pi^0}$ effect on $\Gamma_{\rho}$
                        & \multicolumn{2}{c}{$-0.22$}\\
$m_{\rho^\pm}-m_{\rho^0_{\rm bare}}$ 
                        & $+0.08\pm0.08$ & $ +0.09\pm0.08$ \\
$\pi\pi\gamma$, electrom. decays
                        & $+0.34\pm0.03$ & $+0.37\pm 0.04$ \\
\noalign{\smallskip}\hline\noalign{\smallskip}
                   & $+0.69\pm0.19$ & $+0.72\pm 0.19$ \\
\rs{Total}                   & \multicolumn{2}{c}{$+0.69\pm0.22$} \\
\noalign{\smallskip}\hline
\end{tabularx}
\end{table}
\begin{table*}[htb]
  \caption[.]{\label{tab:br_ee}
    Evaluated \BCVCppz ($\times10^{-2}$)
    from the $\ee$ data including and excluding KLOE data, respectively.
    The errors correspond to the experimental
    uncertainties with the statistical and systematical errors added in
    quadrature (but shown separately for individual experiments).
    The IB uncertainty of $0.22$ is not explicitly quoted for 
    the subcontributions.}
\setlength{\tabcolsep}{0.0pc}
\begin{tabularx}{\textwidth}{@{\extracolsep{\fill}}cccc} 
\hline\noalign{\smallskip}
                          &                 &  \mc{2}{c}{\BCVCppz (\%)} \\
  \rs{Energy range (GeV)} & \rs{Experiment} &  Incl. KLOE & Excl. KLOE \\
\noalign{\smallskip}\hline\noalign{\smallskip}
$m_{\pi^-}+m_{\pi^0}-0.36$ & Combined $\ee$ (fit) & \mc{2}{c}{$0.03\pm 0.00_{\rm exp}$} \\
\noalign{\smallskip}\hline\noalign{\smallskip}
$0.36-0.63$   & Combined $\ee$ & $1.96\pm 0.03_{\rm exp}$ & $1.94\pm 0.03_{\rm exp}$ \\ 
\noalign{\smallskip}\hline\noalign{\smallskip}
$0.63-0.958$ & CMD2 03  & \mc{2}{c}{$20.67\pm 0.13_{\rm stat}\pm 0.12_{\rm syst}$}\\
             & CMD2 06  & \mc{2}{c}{$20.58\pm 0.08_{\rm stat}\pm 0.16_{\rm syst}$}\\
             & SND  06  & \mc{2}{c}{$20.54\pm 0.07_{\rm stat}\pm 0.27_{\rm syst}$}\\
             & KLOE 08  & \mc{2}{c}{$20.26\pm 0.02_{\rm stat}\pm 0.17_{\rm syst}$}\\
             & Combined $\ee$ & $20.40\pm 0.14_{\rm exp}$ & $20.56\pm 0.17_{\rm exp}$ \\
\noalign{\smallskip}\hline\noalign{\smallskip}
$0.958-m_{\tau}$  & Combined $\ee$ & \mc{2}{c}{$2.39\pm 0.06_{\rm exp}$} \\
\noalign{\smallskip}\hline\noalign{\smallskip}
Total        & Combined $\ee$ & $24.78\pm 0.17_{\rm exp}\pm 0.22_{\rm IB}$ & $24.92\pm 0.21_{\rm exp}\pm 0.22_{\rm IB}$ \\
\noalign{\smallskip}\hline\noalign{\smallskip}
\end{tabularx}
\end{table*}
The CVC relation~(\ref{eq:cvc}) allows one to predict the branching fraction 
of a heavy lepton decaying into a $G$-parity even hadronic final state, 
$X^-$, via the vector current
\beqns
   \BR^{\rm CVC}_X=  \frac{3}{2}
                          \frac{\BR_e|V_{ud}|^2}{\pi \alpha^2 m^2_\tau}
                          \int^{m^2_\tau}_{s_{\rm min}}\!\!\! ds\,s\, \sigma^I_{X^0}
                          \left(1-\frac{s}{m^2_\tau}\right)^{\!\!2}
                          \left(1+\frac{2s}{m^2_\tau}\right)\,,
\eeqns
with $s_{\rm min}$ being the threshold of the invariant mass-squared of 
the final state $X^0$ in \ee annihilation. 
This relation was tested ever since the discovery of the 
\Tau lepton. In the best known vector channel, the \ppz final state, 
it has attained a precision of better than 1\%~\cite{md_tau06}, 
and a discrepancy between \BCVCppz and 
\Btau at a level of $4.5\sigma$ was observed.\footnote
{
   The use of the term standard deviation ($\sigma$) in this context requires 
   caution because the results discussed in this paper are mostly dominated 
   by systematic uncertainties with questionable statistical properties.
}
CVC comparisons of \Tau branching fractions are of special interest 
because they are essentially insensitive to the shape of 
the \Tau spectral function, hence avoiding experimental difficulties, 
such as the mass dependence of the \piz detection efficiency and feed-through,
and biases from the unfolding of the raw mass distribution from acceptance and 
resolution effects. 

\begin{figure}[htb]
\includegraphics[width=\columnwidth]{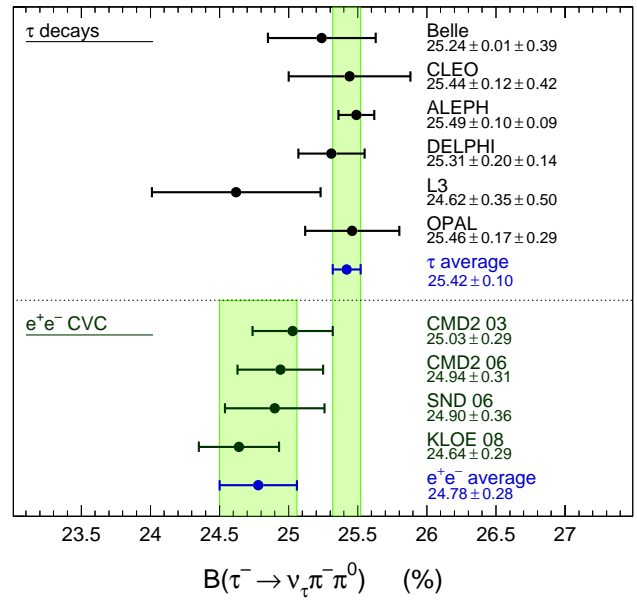}
\caption{The measured branching fractions for 
        $\tau^-\to\pi^-\pi^0\nu_\tau$~\cite{aleph_new,cleo,opal,belle,l3,delphi} 
        compared to the predictions from the $e^+e^-\to\pi^+\pi^-$ 
        spectral functions, applying the isospin-breaking corrections 
        discussed in Sec.~\ref{sec:ib}. 
        For the \ee results, we have used only the data from the 
        indicated experiments in $0.63-0.958\gev$ 
        and the combined \ee data in the remaining energy 
        domains below $m_\tau$. The long and short vertical error bands 
        correspond to the $\tau$ and \ee averages of $(25.42\pm 0.10)\%$ and 
        $(24.78\pm 0.28)\%$, respectively.}
\label{fig:brpipi0}
\end{figure}
Similar to $\Delta a^{\rm had, LO}_\mu[\pi\pi, \tau]$, we have evaluated
the IB corrections to 
\beqn
\Delta \BCVCppz &=& \frac{3}{2}
                         \frac{\BR_e|V_{ud}|^2}{\pi\alpha^2m^2_\tau}
                         \int^{m^2_\tau}_{s_{\rm min}}
                         ds\,s\,\sigma^0_{\pi^+\pi^-}(s)\\
    && \times\;
       \left(1-\frac{s}{m^2_\tau}\right)^2
       \left(1+\frac{2s}{m^2_\tau}\right)
       \left[\frac{\Sew}{R_{\rm IB}}-1\right]\,,\nonumber
\eeqn
where $s_{\rm min}=(m_{\pi^-}+m_{\pi^0})^2$. 
The results are summarised in Table~\ref{tab:ib_br}.
The corresponding $\BCVCppz$ (Table~\ref{tab:br_ee}) is 
$(24.78 \pm 0.17_{\rm exp} \pm 0.22_{\rm IB})\%$ and
$(24.92 \pm 0.21_{\rm exp} \pm 0.22_{\rm IB})\%$, 
based on the combined $\ee$ data, including and excluding the KLOE data, 
respectively. The first error quoted corresponds to the experimental error
and the second error due to uncertainties in the isospin-breaking corrections. 
It differs from the \Tau measurement by 
$(0.64\pm 0.10_\tau\pm 0.28_{ee})\%$ and $(0.50\pm 0.10_\tau\pm 0.30_{ee})\%$, 
respectively, which is still substantial, but less significant than the 
previous result~\cite{dehz03,md_tau06}. 
A graphical comparison between the IB-corrected 
$\BR_{\pi\pi^0}^{\rm CVC}$ and the measured branching fractions 
$\tau^-\to\pi^-\pi^0\nu_\tau$~\cite{aleph_new,cleo,opal,belle} is
shown in Fig.~\ref{fig:brpipi0}. 
The  $\BR_{\pi\pi^0}^{\rm CVC}$ results are obtained
using the \ee data from CMD2, SND and KLOE in 0.63--$0.958\,{\rm GeV}$ and
the combined \ee data in the other energy regions.

\section{Summary}
We have revisited and updated the isospin-breaking 
corrections to \Tau data in the $2\pi$ mode, incorporating new ingredients in 
the long-distance radiative corrections and in the mass and width splittings 
of mesons that enter the pion form factors. We find that the $\tau$ and \ee 
spectral functions from CMD-2 and SND are now marginally consistent, while 
a disagreement with the KLOE measurement remains. The corrected \Tau-based 
result for the Standard Model prediction of the muon $g-2$ is now 
$1.9$ standard deviations lower than the direct measurement, coming closer to 
the \ee value. Similarly, the prediction of the \tautoppz branching fraction 
with \ee annihilation data exhibits a reduced discrepancy with the \Tau measurement.

\begin{details}
~~ 
We are indebted to H.~Hayashii for providing the correlation
of the $\tau$ mass spectrum of the Belle data.
We thank R.~Barbieri, C.~Bouchiat, G.V.~Fedotovich, 
E.A.~Kuraev, W.~Marciano and A.~Vainshtein for helpful discussions. 
GLC and GTS acknowledges Conacyt (Mexico) for financial support.
This work is supported by National Natural Science Foundation of 
China (10491303, 10825524, 10775142), 100 Talents Program of CAS (U-25) 
and the Talent Team Program of CAS (KJCX2-YW-N45).
\end{details}


\begin{thebibliography}{99}
\bibitem{hvptools}   The HVPTools source code (C++, relying on ROOT functionality) and 
                     database (XML format) can be made publicly available. Please contact 
                     the authors.
\end{thebibliography}
\end{document}